\documentclass[a4paper,fleqn]{cas-sc}

\usepackage[numbers]{natbib}

\def\tsc#1{\csdef{#1}{\textsc{\lowercase{#1}}\xspace}}
\tsc{WGM}
\tsc{QE}
\tsc{EP}
\tsc{PMS}
\tsc{BEC}
\tsc{DE}

\begin{document}
\let\WriteBookmarks\relax
\def\floatpagepagefraction{1}
\def\textpagefraction{.001}

\shorttitle{Maritime Cybersecurity: Assessing Health and Safety Risks}

\shortauthors{M Ammar \& IA Khan.}

\title [mode = title]{Cyber Attacks on Maritime Assets and their Impacts on Health and Safety Aboard: A Holistic View}                      


%

\author[1]{Mohammad Ammar}[style=chinese]

\ead{ammargk8497@gmail.com}

\affiliation[1]{organization={Dept. of Computer Engineering, ZHCET, AMU},
    city={Aligarh},
    postcode={202002}, 
    state={UP},
    country={India}}
    
\author[2]{Irfan Ahmad Khan}[type=editor,
                        auid=000,bioid=1]

\cormark[1]


\ead{irfankhan@tamu.edu}



\affiliation[2]{organization={Marine Engineering Technology Department in a joint affiliate appointment with Electrical \& Computer Engineering and Computer Science \& Engineering Department, Texas A\&M University}, 
    city={Galveston},
    postcode={TX 77573}, 
    country={USA}}



\cortext[cor1]{Corresponding author}



\begin{abstract}
There has been an unprecedented digitization drive in the industrial sector, especially in the maritime industry. The profusion of intelligent electronic devices and IOT-enabled cyber-physical systems (CPS) has helped in the efficient use of resources and increased convenience. CPS has enabled real-time remote command and control of industrial assets. Unlike the relatively isolated legacy systems, the intertwined nature of Information Technology(IT) and Operations Technology(OT) brought by Industry 4.0 has increased the complexity of the systems, thereby increasing the attack surface. This work explores the possible consequences of these attacks from a more holistic view, focusing on high-risk assets such as offshore oil rigs, offshore wind farms, and autonomous vessels. The attacks have become more aggressive with the proliferation of such technologies, disrupting the physical process, causing fire and explosion hazards, and endangering human life and environmental health. The possible attack scenarios, the attack vectors, and their physical consequences have been discussed from the perspective of personnel safety and health, along with known security breaches of such nature. To the best of the authors' knowledge, seldom has any work been done that accentuates the possible human and environmental impacts of such attacks.
\end{abstract}


\begin{highlights}
\item The Cyber-attacks on maritime infrastructures are growing rapidly, targeting production, transportation and distribution networks.
\item The profusion of OT and IoT based systems paired with legacy equipment exacerbate the physical effects of cyber-attacks, opening a new era of cyber-terrorism.
\item This study explores the impacts of cyber-attacks in terms of physical damage to the assets, risks to human life and environmental health. 
\end{highlights}

\begin{keywords}
Cyber-Attacks \sep Cyber-Physical \sep Systems \sep Industry 4.0 \sep IoT \sep Maritime Assets \sep Personnel Safety
\end{keywords}

\maketitle
\section{Introduction}
In recent years, the digitization drive has been unprecedented, replacing legacy systems with smart devices, interconnecting almost everything with the internet, from massive data centers to personal wearables. Energy infrastructure is not alien to these developments; it too, has gotten interconnected. The use of smart devices such as Intelligent Electronic Devices (IEDs), IOT-enabled and Cyber-Physical systems, and the adoption of Industry 4.0, in general, has enabled efficient use of resources, less human intervention and increased convenience\cite{4.0}\cite{su132212506}.  Maritime assets, such as ships, containers, port infrastructure, offshore wind farms, and oil rigs, can be equipped with sensors and connected to the Internet, enabling real-time data collection and monitoring. IoT enables remote tracking of assets, predictive maintenance, and improved operational efficiency. CPS enables remote real-time command and control of on-site systems and equipments.  Ships integrated with IoT devices can be interconnected with each other and the internet, sending and receiving information like location, system status, weather, and ocean currrents\cite{article}. Autonomous vessels and smart port utilities like automated quay cranes, automated guided vehicles (AGVs) for container movement, and automated gate systems for efficient cargo clearance use technologies like machine learning, computer vision, and advanced navigation systems employing less human intervention and increased cost-effectiveness\cite{gattuso2023perspectives}. 
\\Similarly, Offshore Wind Farms employ IoT-enabled sensors to monitor and track real-time data across different parameters. Real-time data from these sensors are transmitted to monitoring systems, allowing intelligent systems or manual operators to continuously monitor the performance of wind turbines, discern faults and anomalies, and take appropriate actions accordingly\cite{cite-key} thereby minimizing on-site inspection. The offshore Oil and Gas Industry is embracing the Industry 4.0 paradigm across its multi-faceted assets, which include offshore production facilities, drill ships, storage and offloading units, and complex control and monitoring systems. The integration of Industry 4.0 technologies in these assets brings numerous benefits, including increased efficiency, improved safety, and enhanced decision-making\cite{9096302}.
\\Of all the greatness which embodies in embracing these smart solutions, there is this huge vulnerability to attacks and disruption by adversaries. As the system becomes more digitized and interconnected, the complexity of the system increases, and so does the threat from cyber-attacks and consequently the potential human and financial risks. Oil and Gas Industries are fairly acquainted with cyber attacks for the past 30 years. There has been a 400\% increase in cyber attacks in the Oil and Gas industry since February 2020\cite{cyber}. The February 2020 timeline coincides with the covid 19 lockdown and restrictions as it increased the use of remotely operated and monitoring technologies. Some of the attacks include disruption of offshore operational technology (OT) platforms resulting in the failure of operational capability and safety measures. False data injection in the critical monitoring and controlling facilities, severing of vital communication systems, and sabotaging critical safety infrastructure like gas detectors, flame detectors, and safety triggering mechanisms can entail catastrophic human and financial costs. 
\\Like any other interconnected and digitally-driven infrastructure, offshore wind farms face cybersecurity risks that can have significant consequences. Attackers can exploit the vulnerabilities in the supervisory control and data acquisition (SCADA) system and can trip multiple wind turbines, thus undermining the integrity of the whole system\cite{7406763}. Offshore wind farms can be susceptible to malware and ransomware attacks, where malicious software is introduced into the system, often through phishing emails or compromised software updates. Once infected, the malware can disrupt operations, encrypt data, or demand ransom payments for the restoration of systems. These attacks can cause operational downtime, financial losses, and reputation damage. Similarly, autonomous shipping vessels pose a serious security threat, this potential threat is exacerbated when combined with other maritime infrastructures and shipping vessels. Unauthorized access to the vessel's control system can have devastating consequences, as the intruder can steer the vessel off-course, which can hit other stationary and moving maritime assets, using it as a battering ram. This will not only cause financial costs but can entail serious human loss. \\
There are very few cases of these attacks being publicized. Most of it, if not all, stays confidential as it may undermine the reputation at both the business and state levels\cite{negpublicity}\cite{kechagias2022digital}. Further, most studies in the literature about cyberattacks on marine infrastructure consider its societal, financial, and supply chain loss impacts, among others. To the best of the authors' knowledge, only a few studies have been performed on the impacts of cyberattacks on the health and safety of the personnel working on marine infrastructure. Therefore, the purpose of this study is to identify and explore the less studied possible human risks associated with the automation and digitization of these maritime assets. Also, it will dig into the impacts and consequences of cyberattacks on the health and safety, human life, and environmental health  of personnel and marine infrastructure. 
\begin{figure}[htb!]
  \centering
  
  \includegraphics[width = 0.5\textwidth]{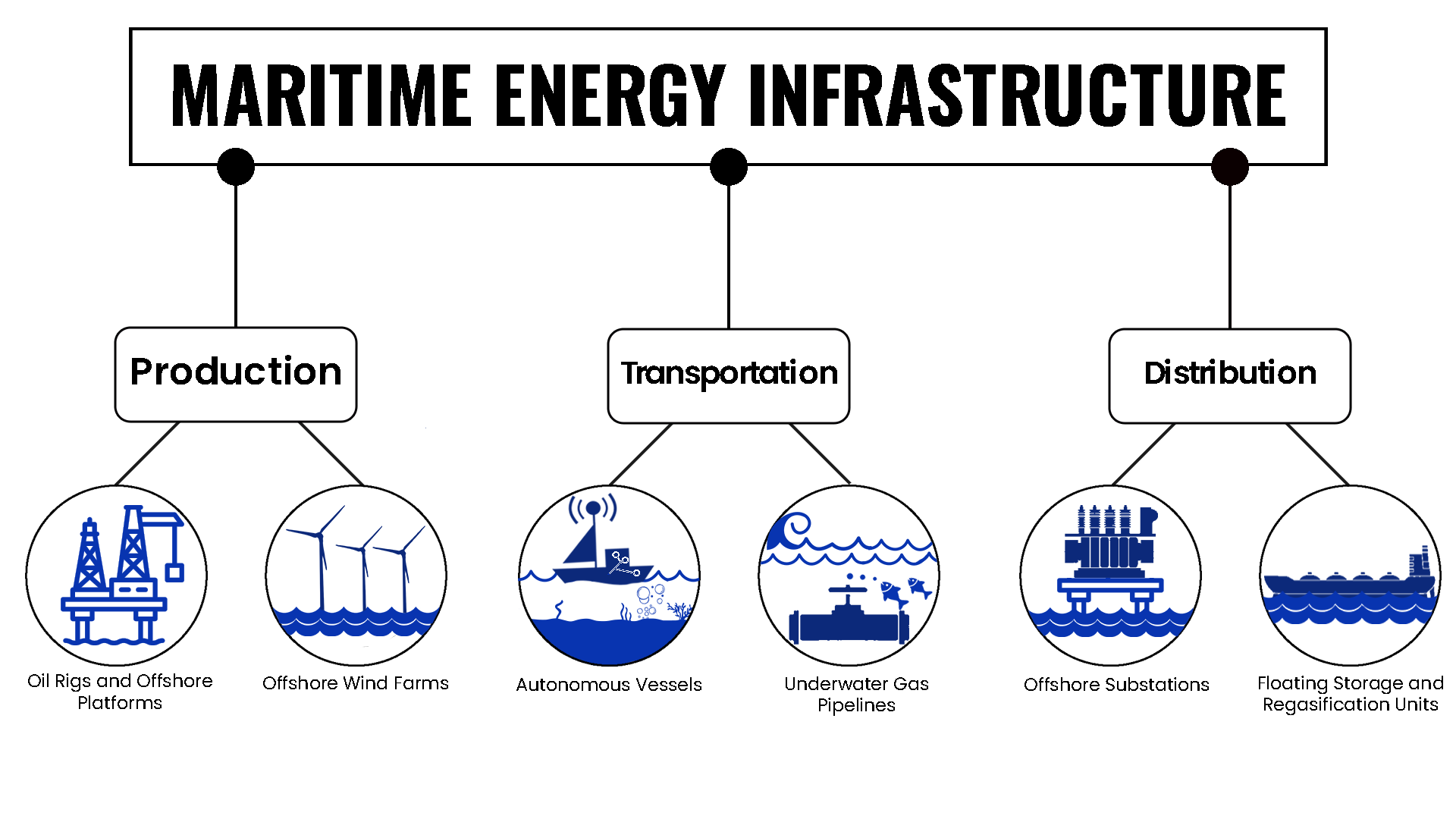}
  \caption{Division of Maritime Assets}
  \label{fig:RSUencountered}
\end{figure}

The rest of the paper is structured as follows: In the following section, we shall get acquainted with the maritime infrastructure covering Offshore Wind Farms, Upstream Oil and Gas Infrastructure, Smart ports, and Autonomous Shipping Vessels. In section III, we shall briefly explain possible types of cyber attacks on maritime infrastructure along with some previous known attacks on CPS. In section IV, we will hypothesize some possible attack scenarios and evaluate their possible human and health costs and finally, section V will conclude the paper. 

 \section{Maritime Energy Infrastructure}
Maritime energy infrastructure constitutes multifaceted assets that can be broadly classified into three categories 1). Production 2). Transportation
and 3).
Distribution. Although Ports and Autonomous Shipping vessels are not directly considered under the energy infrastructure, they are often used for Transportation and Distribution.\\
\subsection{Production}
\subsubsection{Oil Rigs and Offshore Platforms}
These structures are located in offshore waters used to explore and produce oil and gas resources. Offshore platforms can be fixed structures (such as platforms built on seabed or elevated platforms) or floating structures (such as floating production storage and offloading vessels or floating platforms). These platforms are equipped with drilling equipment, including a drilling rig, drill pipe, blowout preventers, and well control systems.

The drilling process involves drilling a wellbore into the seabed to access oil or gas reservoirs\cite{2005iv}. Software systems are used to monitor and control drilling parameters, such as drilling speed, pressure, and mud circulation.
\begin{figure}[t]
  \centering
  
  \includegraphics[keepaspectratio, width = 0.4\textwidth]{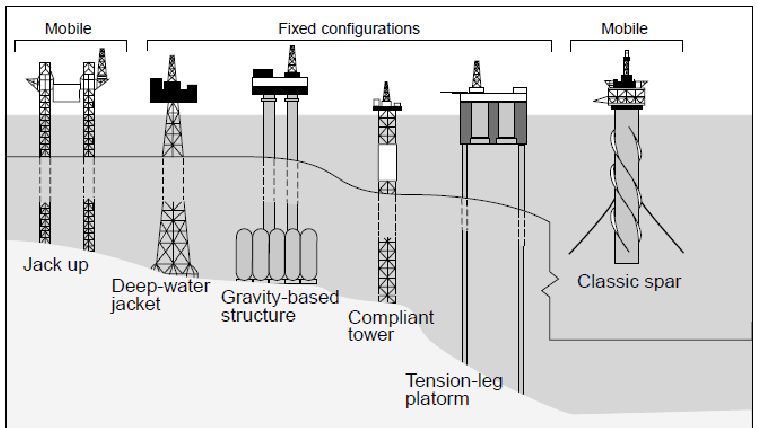}
  \caption{Types of fixed offshore platform (Yew 2014)}
  \label{fig:RSUencountered}
\end{figure}
Once a well is drilled, offshore platforms facilitate the extraction and processing of hydrocarbons. Oil and gas production systems, such as wellheads, production trees, and risers, are used to control the flow of fluids from the wells to the platform. Production processing facilities, including separators, pumps, compressors, and treatment systems, are employed to separate oil, gas, and water and prepare them for transportation. Offshore platforms incorporate sophisticated Industrial Cyber-Physical Systems (ICPS)  which allows for real-time remote monitoring and control\cite{6461894}\cite{8410404}. Supervisory Control and Data Acquisition (SCADA) systems and Distributed Control Systems (DCS) are used to monitor and control various processes, equipment, and safety systems on the platform. These systems collect data from sensors and instruments distributed throughout the platform and provide operators with real-time information on operational parameters\cite{CARVAJAL201843}. Previously, these Industrial Control Systems were proprietary and were used on local networks\cite{7245330} fairly isolated from the outside world\cite{Bundi2020EFFECTSOC}. However, with recent advancements in Operational Technologies(OT) and the need to make better and faster decisions, these systems are now being integrated with various IT based industrial technologies\cite{9136701}. This has led to a greater integration of these ICS with the outside network. This integration has opened a larger potential attack frontier for the adversaries. Any attack targeting the plant can have serious consequences. Failure of the Blowout Prevention system can result in a sudden and powerful release of hydrocarbons. Blowouts can result in explosions, fires, and the release of toxic substances into the environment, endangering both human workers and marine life. Cramped living quarters can hamper the evacuation procedures, again jeopardizing the rescue operation.

\subsubsection{Off-shore Wind Farms}
Off-shore wind farms consist of multiple wind turbines installed in strategic locations. The turbines are anchored to the seabed using foundation structures, such as monopiles, jacket foundations, or floating platforms. Each wind turbine comprises several components, including the rotor, nacelle, tower, and foundation.
\begin{figure}[ht]
  \centering
  
  \includegraphics[keepaspectratio, width = 0.4\textwidth]{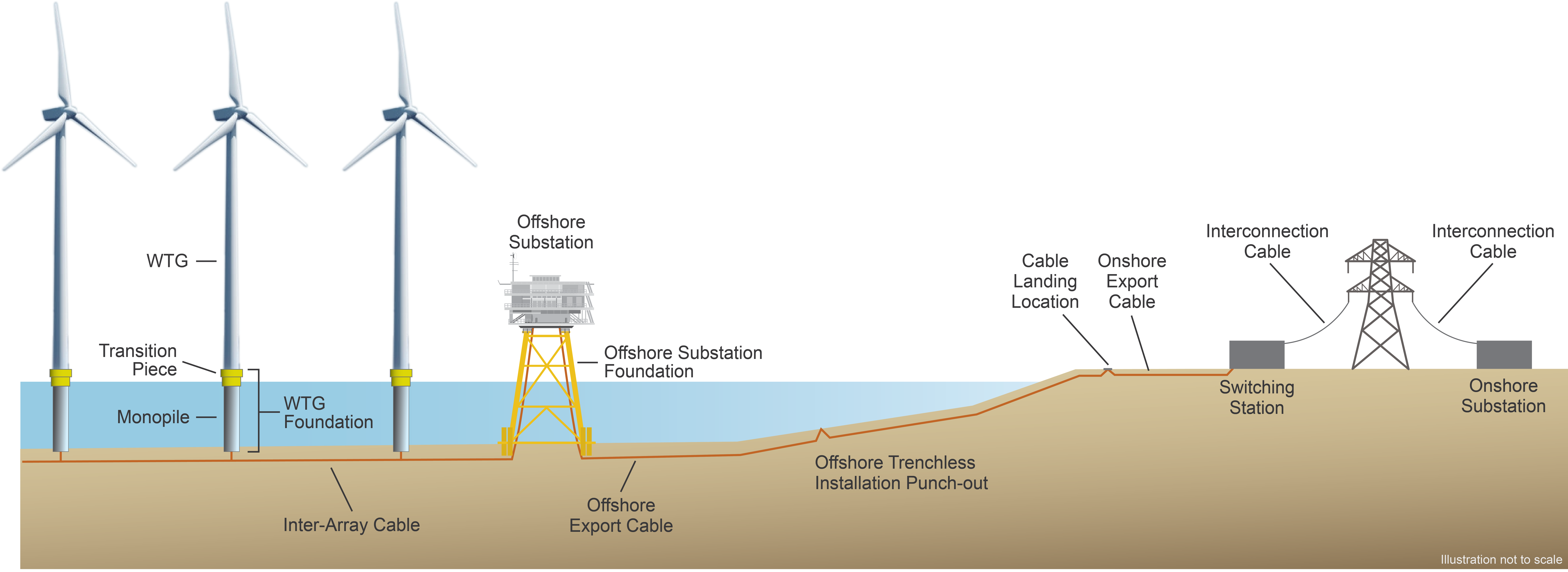}
  \caption{Off-shore Wind farm }
  \label{farm}
\end{figure}
The wind turbines convert the kinetic energy of the wind into mechanical energy through the rotation of their rotor blades.
\begin{figure}[ht]
  \centering
  
  \includegraphics[keepaspectratio, width = 0.4\textwidth]{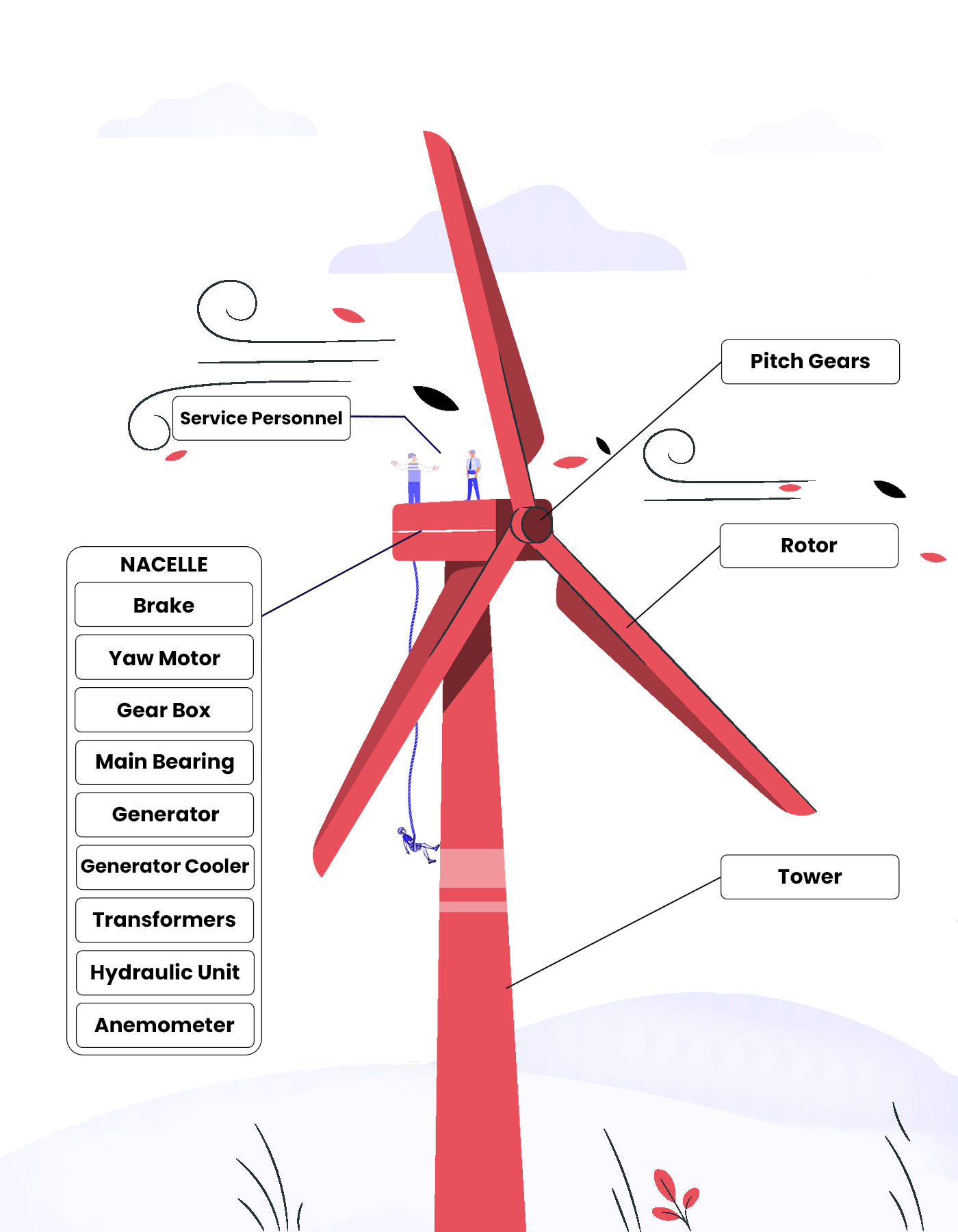}
  \caption{Wind Turbine }
  \label{farm}
\end{figure}
The rotor blades spin a generator within the nacelle, which produces electricity. Power electronics and transformers within the nacelle convert the generated electricity to a suitable voltage for transmission. Offshore wind farms utilize sophisticated communication and control systems to monitor and manage the turbines, substations, and overall farm operations. The task of gathering grid data—measured electrical variables at the wind turbines—is carried out by the wind farm SCADA systems. This data is transmitted to a  control center for monitoring and control purposes\cite{7406763}. The control system can alter various turbine parameters like pitch and yaw of the turbine blades, it can also set the maximum rotor speed and can apply emergency breaks in case of fast winds\cite{STAGGS20173}. In the case of floating wind turbines, the ballast tanks can also be controlled. Vulnerabilities in the Supervisory Control System can result in unauthorized access resulting in the errant behaviour of turbines, which can have potential human and financial risks.

\subsection{Transportation}
\subsubsection{Autonomous Vessels}
Autonomous vessels, also known as unmanned or autonomous ships, are maritime vessels that operate without an onboard crew or with minimal human intervention. These vessels leverage advanced technologies, software systems, and hardware components to navigate, operate, and communicate.
Autonomous vessels rely on a combination of technologies for navigation, situational awareness, and decision-making. They are equipped with various sensors, such as radar, lidar, sonar, GPS, and cameras, to perceive their surroundings and collect real-time data about the environment. Control systems, including autopilot and dynamic positioning systems, enable autonomous vessels to control their propulsion, steering, and maneuvering. These systems use data from sensors, navigation algorithms, and feedback mechanisms to maintain course, avoid collisions, and perform docking operations. The relation of autonomous vessels to maritime energy transportation lies in their potential to improve efficiency, safety, and sustainability in the movement of energy resources. Autonomous vessels can be used for transporting various forms of energy, such as oil, gas, LNG, or renewable energy components, across maritime routes\cite{doi:10.1080/16258312.2019.1631714}. By leveraging autonomous technology, these vessels can optimize navigation, reduce human errors, enhance fuel efficiency, and enable continuous operations.
\subsubsection{Underwater Gas Pipelines}
Underwater gas pipelines are critical infrastructure that enables natural gas transportation across bodies of water, such as oceans, seas, and lakes. These pipelines play a significant role in the maritime energy transportation sector. They are typically made of steel or high-density polyethylene (HDPE) pipes, designed to withstand the harsh marine environment. The pipelines consist of a continuous length of interconnected pipe segments that form a route from the gas source (e.g., offshore gas field) to the receiving terminal onshore or to other interconnecting pipelines. Various software and hardware systems are used to monitor and maintain the integrity of underwater gas pipelines. These systems include:
\begin{itemize}
    \item Pipeline Inspection Gauges (PIGs)
    \item Subsea Sensors
    \item Remote Operated Vehicles (ROVs)
    \item Fiber Optic Systems
    \item Emergency Shutdown Systems (ESD)
\end{itemize}

\subsection{Distribution}
\subsubsection{Offshore Substations}
The foundation of offshore wind farms is offshore substations, commonly called offshore electrical platforms. They act as centers for gathering, converting, and distributing the electricity produced by wind turbines. Large steel constructions are generally used for offshore substations since they are made to endure the severe sea environment. They are made up of numerous levels or decks that can hold different systems and equipment. The design considers stability, buoyancy, and resistance to wave, wind, and ice loads. As with any other substation, the offshore substation uses SCADA systems for monitoring and maintenance. Various relays, circuit breakers, and other safety devices are used to protect the station from electrical faults. 

\subsubsection{Floating Storage and Regasification Units (FSRUs)} 
Liquefied natural gas (LNG) is imported, stored, and regasified using Floating Storage and Regasification Units (FSRUs), which are specialized ships. They offer a versatile and affordable way to receive LNG from LNG carriers, store it onboard, and then transform it back into gas for distribution. FSRUs are anchored near the shore and receive LNGs from the carrier ships in liquid form through cryogenic arms and flexible hoses. The LNG is stored at 109K - 110K. Before distribution, the LNG is converted into gas by heating it using seawater and then pushed into pressurized pipelines.
Due to the inherently hazardous nature of the LNG, FSRUs incorporate advanced safety systems to ensure the secure handling of LNG and to prevent accidents. These systems include fire and gas detection systems, emergency shutdown systems, and containment measures to prevent leakage or spills.

\section{A Background on Cyber Security}
Cybersecurity is the practice of preventing unauthorized access, damage, and theft to computer systems, networks, and data. It involves a variety of safeguards, technologies, and best practices meant to stop cyberattacks and protect information.

\subsection{Types of Cyber-Attacks}
There are many types of cyber attacks, but in the scope of our study, we are most interested in cyber attacks on marine energy systems and their impacts on health and safety. Although cyber-attacks cause great monetary loss, a cyber attack on a marine energy system entails a loss of human life and other health hazards.
\begin{figure}[ht]
  \centering
  
  \includegraphics[keepaspectratio, width = 0.4\textwidth]{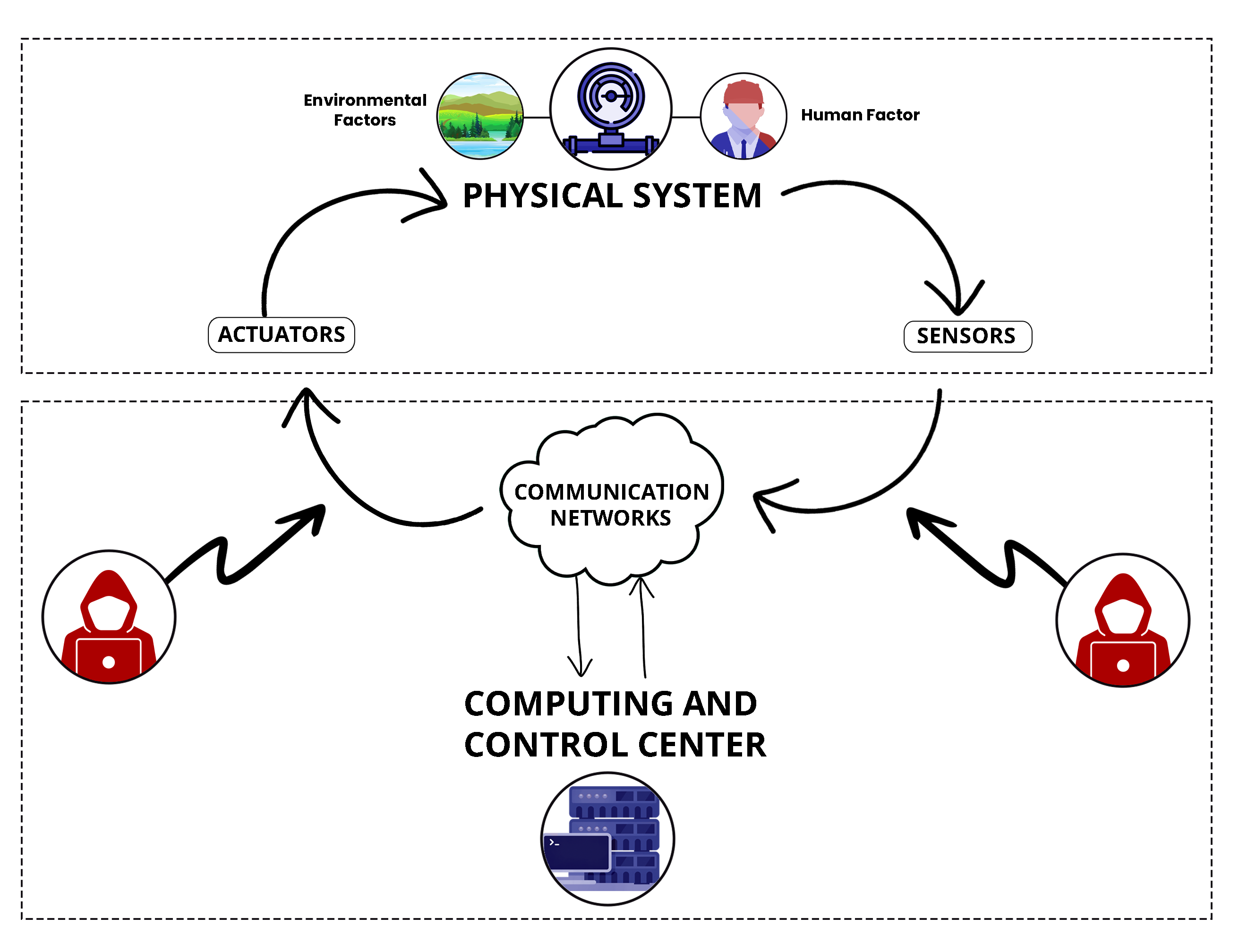}
  \caption{General overview of cyber-physical systems and the attack vectors. }
  \label{farm}
\end{figure}

\subsubsection{Zero-day Attacks}
A zero-day attack on marine energy systems refers to exploiting previously unknown vulnerabilities or weaknesses in the software or hardware that controls physical devices, infrastructure, or industrial systems. These attacks can have severe consequences as they target the integration between digital and physical components, affecting critical infrastructure and industrial operations. Some of the famous Zero-day attacks are Stuxnet Worm (2010); Stuxnet was a sophisticated computer worm that exploited zero-day vulnerabilities in Microsoft Windows and Siemens SCADA (Supervisory Control and Data Acquisition) systems, which were used to control and monitor industrial processes in the nuclear enrichment facility. Stuxnet manipulated the centrifuges' speed control system, causing physical damage to the uranium enrichment process and sabotaging the facility's operations. Another recent one is the Triton/Trisis Malware (2017). It targeted a petrochemical plant in the Middle East, sabotaging the safety instrumented system (SIS). This system was responsible for shutting down the processes in case of an emergency. The malware allowed the attackers to reprogram the system to cause catastrophic physical consequences. Similarly, Maroochy Water Services Attack (2000), while technically it was not considered a Zero-day attack, it serves as an example of how cyber-physical systems can be targeted. Australian hacker named Vitek Boden gained unauthorized access to a sewage treatment plant's control system. He used radio transmissions to manipulate pump controls, causing millions of liters of raw sewage to spill into waterways and parks\cite{10.1007/978-0-387-75462-8_6}. This attack can serve as an analogue scenario to the cataclysm that can be caused on the maritime energy infrastructure by these types of attacks. 
\\

\subsubsection{ Eavesdropping Attacks}
While eavesdropping attacks do not have imminent physical consequences, it does threaten the information integrity of the system. Eavesdropping attacks are typically passive. The attacker does not actively modify or disrupt the communication but listens to the exchanged data. While sounding harmless, this sensitive data can actually be used to stage a large-scale attack on the system.
\\

\subsubsection{Denial of Service Attacks (DoS)}
DoS attacks aim to disrupt the availability of computer systems, networks, or websites by overwhelming them with a flood of requests or traffic\cite{MAHMOUD2019101}. This prevents legitimate users from accessing the targeted resource. In maritime energy systems, DoS can have severe consequences, disrupting the integration between digital and physical components. There are many different types of DoS attacks, such as Ping Flood Attack, SYN Flood Attack, HTTP/HTTPS Flood Attack, NTP Amplification Attack, DNS Amplification Attack, and Slowloris Attack. In the Industrial Control System (ICS) domain, these attacks can prohibit the legitimate user from enforcing countermeasures to protect the machinery from erratic behavior. In case of an unforgiving marine environment, these failures can result in potential harm and life-threatening consequences for the on-site crew. 

\subsubsection{Data Injection Attack}
Data Injection Attack is also known as Data Spoofing Attack or Falsification Attack. In these types of attacks, the communication channel is compromised, whereby the attacker can manipulate or insert false data into a system or database. This is done by exploiting the vulnerabilities in the communication protocol, software, and device itself. The attacker can go through the physical layer using frequencies to manipulate the communication or through the cyber-layer, assuming he has access to the credentials\cite{8396855}. In both cases, altering the signals sent to the actuators and disrupting the nominal state of the plant. FDI can also target the sensor reading of the plant, thereby deceiving the system or the remote operator. Although there are measures to counter these attacks, sophisticated and highly coordinated attacks can still bluff the security measures. Mo \textit{et al.}\cite{mo2010false} provides optimal conditions under which the attacker can bypass the failure detection systems, Zhang \textit{el al.}\cite{zhang2020false} proposed a stealthy FDI attack which is robust to residual-based detectors. 

\subsection{Some Notable Cyber- Attacks on CPS}

There are few attacks that are in the public domain, and most, if not all, are confidential. One of the most notable attacks, as mentioned earlier, was the Maroochy Water Service attack. A disgruntled employee manipulated the control pumps and was able to release large amounts of sewage, flooding the parks and waterways\cite{10.1007/978-0-387-75462-8_6}\cite{oueslati2019comparative}. The Stuxnet attack of 2010 targeted Iran's uranium enrichment facility. The malware exploited four zero-day vulnerabilities to sabotage the centrifuges' nominal operations, thereby halting the enrichment process\cite{5772960}\cite{baezner2017stuxnet}. The Stuxnet worm infected the facility computers through a USB stick and was able to infect the windows operating systems and siemens step7 software used to program industrial control system. The worm caused the centrifuge to spin above its nominal speed and injected false data into the supervisory systems causing the operators to be oblivious to the attack.  In 2014, the German Steel Mill's network was compromised, and the attackers were able to disrupt the shutting down process of the blast furnace, causing widespread destruction to the facility\cite{lee2014german}. A similar incident occurred in the U.K. where two people lost their lives, and 13 were injured. However, this was caused by a human error instead of a cyber attack at the German Plant. These two events serve as a testament to the dangers posed by compromised Cyber-Physical systems\cite{miller2021looking}. The Ukrainian power grid attack of 2015 caused a blackout across Ukraine, affecting around 200K civilians. The attack targeted the SCADA systems, which controlled the distribution substations\cite{case2016analysis}. One of the most recent attacks was on Oldsmar's water treatment facility. The attacker gained access to one of the consoles of the facility using a compromised credential of a publicly available remote access application namely TeamViewer. The attacker logged on to the console and, using the Human Machine Interface(HMI) increased the concentration of Sodium Hydroxide, also known as Lye, from 100ppm to  11100ppm, a hundred times increase in the concentration\cite{cervini2022don}. Lye in these amounts can cause severe burns, permanent damage to the organs, and even death\cite{atug2009critical}. This attack coincided with the Super Bowl event in the area. Thankfully, the attack was foiled. If it had been successful, thousands of people would have been affected. This attack serves as a wake-up call for the government as well as the industry at large. 
\begin{table}[htbp]
    \centering
    \caption{Some of the Cyber-Attacks from 2000 Onwards}
  \begin{tabular}{p{0.1\columnwidth}p{0.2\columnwidth}p{0.5\columnwidth}}
        \hline{} 
                Year & Country & Details \\
         \hline{} 
                2000 & Australia & Causing sewage to spill in waterways \\        
        \hline{} 
                2010 & Iran &  Stuxnet attack destroying core controllers of the industries  \\
         \hline{} 
                2014 & Germany & Attacks led to the disruption of the shutting down process of blast furnace \\        
        \hline{} 
                2015 & Ukraine &   BlackEnergy attack on power grid leading to massive power outage  \\
        \hline{} 
                2017 &  Russia, Ukraine India, China  & WannaCry attack aiming to encrypt data and demand ransom payments  \\
        \hline{} 
                2020 &  Brno University Hospital Czech Republic  &   A    cyber attack that shut down IT  network of a Czech hospital  \\
        \hline 
                2020 &  US Dept. Health   Human Services  & Unspecified attack on servers \\
        \hline 
                2021 & Colonial Pipeline, US &  A ransomware attack on a US fuel pipeline, leading to shutdown of a  critical fuel network  \\
         \hline{} 
                2021 & United States & Attacker gains access to the water treatment facility's systems and increased the concentration of NaOH in the supply water \\        
        \hline
    \end{tabular}
\end{table}
\section{Potential Hazard  to Humans and Environment in Maritime Infrastructure}
The proliferation of cyber physical systems and IoT based control and monitoring technology, in the maritime industry has made it highly susceptible to cyber physical attacks. The complexity and hazardous operational nature coupled with the high asset value of the infrastructure all together attract perpetrators to target these systems. Due to the socio-technical nature of the system, any attack targeted at the physical process can have serious consequences, Coupled with the remote and unforgiving environment of the sea, these attacks can have life threating outcomes. In the upcoming sections, we shall discern the potential effects on human life of these attacks in each of the maritime infrastructures.  
\subsection{Offshore Oil \& Gas}
\begin{figure}[h!]
  \centering
  \includegraphics[width = 0.8\textwidth]{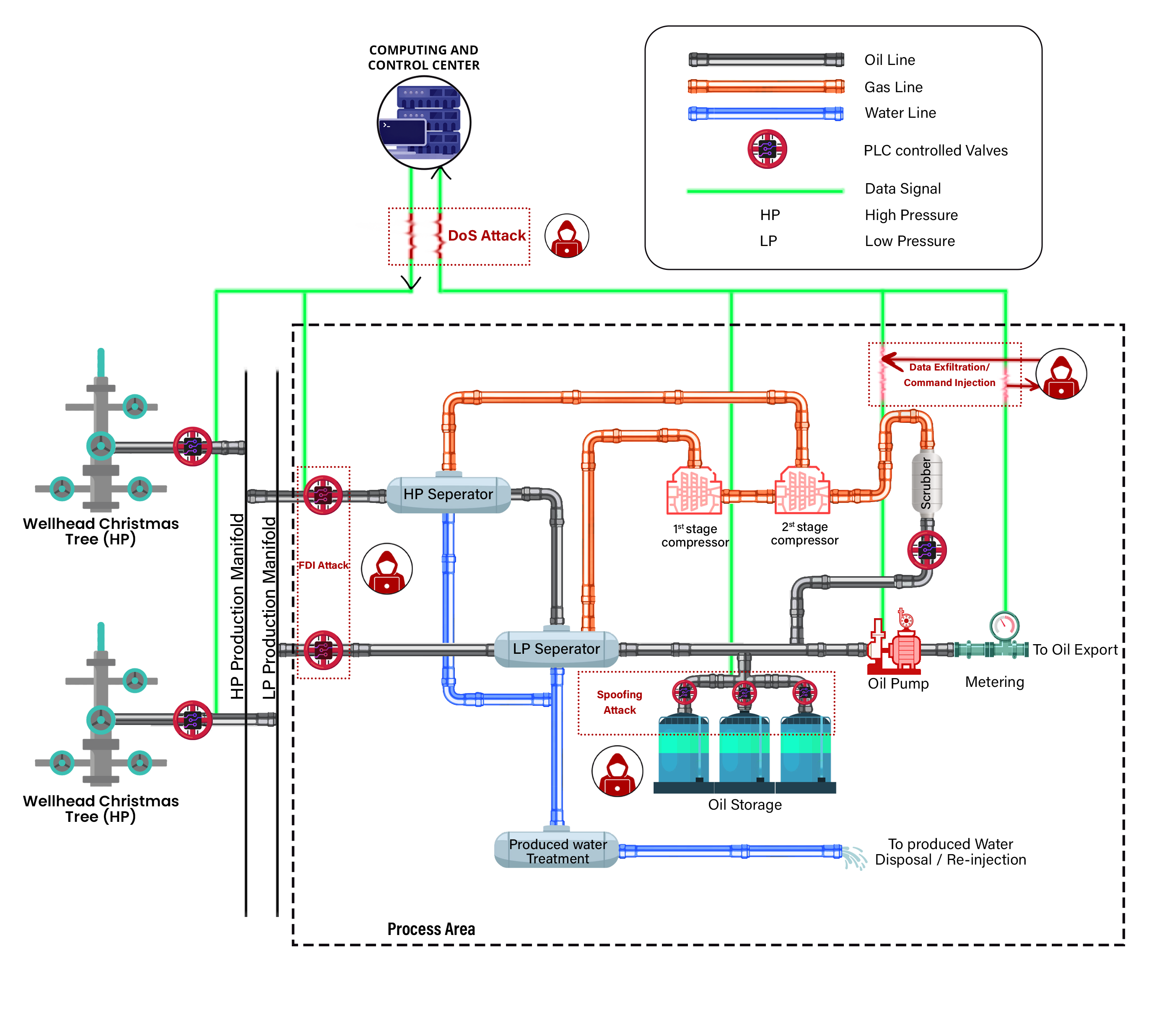} 
  \caption{Schematic Representation of process facility and possible attack vectors. }
  \label{fig:processarea}
\end{figure}

The Offshore Oil Facility is composed of a drilling rig, transportation system, processing plant, accommodation and utilities for staff, and the structure itself. The process plant is where the production of oil and gas takes place before transporting it to the onshore facilities. The processing plant has several levels of processing units which include separation, compression, and dehydration as shown in fig \ref{fig:processarea}. 
There are quite a few types of risks associated with the Oil and Gas facilities but we shall only focus on the risks associated with the process itself.  The process facility at the oil rig is a small dense chamber with many machines and equipment. These areas are overcrowded with compressors, separators, and high-pressure pipelines carrying highly flammable materials. Any mishap can have catastrophic consequences and can also lead to a domino effect. To illustrate the hazard of cyber attacks on these facilities a simple example is adapted from Khakzad \textit{et al.}\cite{khakzad2011safety}. A propane injecting control system feeds a scrubber with propane from a gas pipeline. To maintain the required amount of pressure in the scrubbing chamber the pipeline is equipped with an actuator operated automatic valve. In the Industry 4.0 era it is plausible to assume it as a cyber-physical system connected over the internet. The control system has an actuator, a pressure transmitter, and a pressure controller. A MitM attack if successfully executed can stealthily alter the values of the pressure transmitter. The insertion of malicious readings will cause the pressure to increase in the scrubbing chamber. The altered readings displayed on the monitoring station will cause the entire system to be oblivious to the attack, hence forestalling any remedial measures.
\par Taking forward the analogy a similar incident is comprehensible at the offshore O\&G  platforms. As mentioned earlier, the processing facility houses high-pressure units and flammable materials. The fluid from the well is passed through the first separator which separates crude oil, water, and gas followed by a second separator which separates gas condensate and water. This is followed by compressor units, a flash drum, and a drier. The pressure in the separator is controlled by the flow valve. A PLC command injection attack targeting the flow valve controller can cause the pressure to increase in the separator. This increase in pressure if not mitigated can cause the Separator to fail as Boiling Liquid Expanding Vapour Explosion (\textbf{BLEVE}).  These explosions produce a fireball, whose emissive power typically ranges from 180kW/$m^2$ \cite{opschoor1979methods}to 350kW/$m^2$\cite{kumar1996guidelines}. The shock wave velocity is typically 753m/s with sub-sonic fragment velocity. The fragments resulting from the explosion are hard to simulate, but many researchers have used deterministic models and stochastic simulations to determine its 
characteristics\cite{gubinelli2004simplified}\cite{hauptmanns2001procedure}\cite{gubinelli2009assessment}. The direct exposure to the fireball can cause severe burns, while the shock wave and peak overpressure can cause barotrauma, resulting in tympanic membrane rupture and middle ear damage\cite{fadaei2015numerical}, injury to the lung parenchyma, injury to brain parenchyma and injury causing abdominal hemorrhage and perforation. These are categorized as the primary injuries caused by the explosion\cite{tan2015fast}. 
\par The secondary injuries include those which are caused by the fragments and projectiles like of metals and concrete. The tertiary category accounts for injuries caused by human transposition and impacts, like hitting a wall or furniture by being thrown away. Injuries can include fractures, amputation, and multiple types of soft tissue injuries. Quaternary injuries result from the toxin released by the explosion and other effects which are not covered by the above three categories. The effect of BLEVE will not stop here as it may cause a domino effect, critically damaging other infrastructure and causing fires\cite{kumar2014study}. Other critical process plants can also be affected in a similar way, either by injecting erroneous values in PLC controllers, by spoofing sensor readings or sabotaging the automated safety mechanism by changing the safety values in the PLC controllers or the manual intervention by denial of service attacks (DoS). Table 2 summarises some of these attacks and their impact. All in all, it is of utmost importance to identify these vulnerabilities, quantify the threat probabilities, and to mitigate the risks. 

\begin{table*}[t]
    \centering
    \caption{Some of the Cyber-Attacks in O\&G process facilities and its impacts}
  \begin{tabular}{p{0.10\textwidth}p{0.20\textwidth}p{0.20\textwidth}p{0.20\textwidth}p{0.20\textwidth}}
        \hline{} 
                Process & Attack & Target Component & Consequence & Environment \& Human Impact  \\
         \hline{} 
                High Pressure Separator & FDI, Malicious Command Injection & Flow Valve PLC controller, Pressure and Temp Sensor & Over-Pressure, BLEVE & Burns, Damage to Air-filled organs, penetration by projectile, pool of fire \\        
        \hline{} 
                Emergency Shutdown and Disaster Mitigation Systems &  DoS attack & PLC controller, actuators, Safety Instrumented System  & Operation Under Undesirable Conditions & Damage to Equipement, Possibility of Spillage, Potential Loss of Marine or Human Life \\  
         \hline{} 
                Low Pressure Separator & FDI, Malicious Command Injection & PLC controller, Pressure and Temp Sensor & Over-Pressure, BLEVE & Burns due to fireball explosion, Release of Toxin, pool of fire \\    
        \hline{} 
               Compressor Unit & FDI, Malicious Command Injection & PLC controller, Pressure and Temp Sensor & Over-Pressure, Leakage & Jet Stream of fire causing burns and potential loss of life \\  
        \hline{} 
                Storage Unit & Spoofing Attack & Level Sensor & Spillage or Tank Rupture & Burns due to fire, Oil Spill affecting marine life. \\  
       
        \hline
    \end{tabular}
\end{table*}

\subsection{Offshore Wind Farm}
The offshore wind farms(OWF) are a cluster of wind turbines located off the shore, these structures harness energy from the wind that blows over the sea. The collected energy is transmitted over to the substation either offshore or onshore which is then further distributed. Unlike the oil rig structures which has a permanent human presence, these structures are unattended. Most of the systems are automated and the monitoring is done through software like SCADA. However maintenance is done physically, this may involve climbing the nacelle or grappling down the turbine itself. The sheer scale of the wind farm and the remoteness of it makes any rescue attempt futile. The OWF employs a flat logical control structure with multiple turbines connected in a ring topology. An array of sensors are used to gather data on the status of the turbines, these information is relayed to the control station which uses a combination of IT and OT to control various parameters to optimize the energy output. \ref{farm} shows the structure of wind turbines, the nacelle holds the major components of the turbine including the generator and gear assembly. Most of the turbine systems are autonomous-- adjusting the blade angles and nacelle direction in order to optimize the energy output. One major component of the system is the braking mechanism, this helps to ensure that the turbine operates under the required RPM. An attack on the PLC controllers can compromise the control system of the turbine. This can give the attacker the privilege to change the direction of the nacelle, the orientation of the blades and the brakes itself. These parameters are vital for the optimization of energy output. However these three control surfaces can be used to push the wind turbine beyond safe operational RPMs. The wind turbines are designed to cut off in high gusty winds or when the RPM reaches beyond a certain threshold. These systems ensure that the turbine experiences mechanical and aerodynamic stresses in acceptable ranges. 
\par An attack on these surfaces can compromise these safety systems. A turbine operating at extreme overload can experience multiple failures. Various publications have studied the failure rate of these systems, mainly focusing on the technical components. From the perspective of personnel safety, \textit{blade flew off}, \textit{partial blade failure} are of the utmost interest. The blade flew off, or the partial blade flew off, is the structural failure of the blade, causing the entire blade to fly off or sometimes in pieces as projectiles. \cite{rogers2012method} provides an in-depth analysis of the throwback region. These high-energy projectiles can cause severe damage to the human body. These can include cuts, lesions, amputations, and even death. Operation of the wind turbines  beyond safe limits can also lead to complete structure failure. In the case of offshore turbines a service vessel is also present. A complete structure collapse can cause the nacelle to fall. The weight of offshore wind turbine nacelle typically ranges from 100 to 600 Ton. An impact of this energy on the vessel deck can be devastating. 
\subsection{Underwater Oil Pipeline}
The sub-sea crude oil pipeline is an important part of oil transportation network along with oil transporter and ships. These pipelines are continuously monitored by sensors and computers to maintain the flow rate and pressure of the oil. Some of these sensors are used to detect oil leakages. A sensor spoofing attack can render the sensor useless, hence keeping the whole system oblivious to the leakage. This can cause large amount of crude oil to leak before detection, causing mayhem in ocean habitats. Attacks can also effect the PLC based PID controllers which are used to control flow rate through SCADA software. These changes in operational parameters can cause erratic pressure and flows in the pipeline causing it to leak or explode. In 2009 a disgruntled employee was found guilty of disabling the computer systems that were designated to detect oil leaks in pipeline off the coast of Southern California\cite{WEBSITE:2}. In another event in 2008 a section of the Baku-Tbilisi-Ceyhan (BTC) pipeline exploded, an investigation into the explosion later by the U.S. shed light on a cyber-attack. The hacker got access to the pipeline monitoring system and shut down the alarm warning system and was able to super pressurise the pipeline causing it to explode\cite{WEBSITE:3}.

\subsection{Autonomous Ships and Unmanned Vessels}
Autonomous Ships and Vessels possess unmatched potential to cause destruction in the maritime infrastructure. Currently, their use is restricted to small and local vessels. However, the trend shows that they will be used in deep-sea voyages and general shipping by 2030\cite{WEBSITE:1}. These vessels can either be completely autonomous or commanded by a shore control center. Relying on multiple sensors and communication protocols for navigation and actuators for control, these systems are vulnerable to attacks as much as any other cyber-industrial system. However, unlike static assets like wind turbines and oil rigs, these assets pose a wider threat to the maritime industry. The attacks on the shipping industry are not new and have been happening for quite a long time. One attack targeted the Electronic Chart Display and Information System (ECDIS), resulting in the loss of critical information\cite{bimco2018world}. Another attack compromised the GPS system, indicating the ship to be 32 km inland\cite{newman2019cyber}. With the advent of autonomous ships, new attack surfaces are to be recognized. A compromised container can be used as a battering ram by the perpetrators-- opening a new kind of terrorism. One scenario could be to use the hijacked vessel to ram it against other stationery assets like wind farms, oil rigs, and other regasification units. Another scenario could be to use these ships to ram other vessels and ports. A vessel colliding with an oil rig can have severe consequences. However, the extent of damage depends on various factors such as vessel size, speed, vessel design, type of oil rig and other factors. \cite{f0e54f24a09d4b7f9a700ec83cf75665} has done a comprehensive analysis of these collisions. Considering the worst-case scenario, a small vessel of 5000 ton at a modest speed of 12 knots possess around 200 MJ of kinetic energy, and most of the offshore structures have kinetic resistance of less than 200 MJ. An impact with such energy can cause the oil platform structure to fail. The most imminent damage will be to the oil riser which can snap under such impact load. This can cause huge amounts of crude oil to be released into the ocean. Just like the Deep Water Horizon, these blowouts can cause the destruction of marine life both on the surface and in the deep waters. Oil spills can affect microbial life such as phytoplanktons as well as fisheries\cite{beyer2016environmental}, the impact on the ecosystem is highly complex to determine and is out of the scope of this study. Apart from the environmental impact, there is also a high fatality risk associated with topside events such as fires and explosions. Secondary fatalities and injuries are from structural impacts such as debris falling and other flying shrapnel.
\section{Conclusion}
This work attempts to accentuate the possible consequences of cyber attacks in the maritime industry on the health and safety of personnel working abroad. In this paper, effort is made to shed light on those consequences, which are often eclipsed by the monetary and economic factors -- namely the human and environmental impacts. We started with a brief description of the trend in the maritime industry, the proliferation of digital technologies, sensors, cyber-physical systems, and the adoption of Industry 4.0 in general. The maritime sector is evolving rapidly, with new technologies being introduced regularly. The industry is ever more susceptible to cyber-attacks. Attacks can now impact the physical systems, causing financial losses and endangering human lives and the environment, causing great destruction. The attack surface will continue to grow as the technology advances, becoming more intertwined and ever more complex, engendering new risks and consequences. At the time of publication, to the best of our knowledge, there is seldom any work that draws attention to the possible human and environmental factors of such attacks. There is an eminent need for the qualitative and quantitative assessment of risks associated with these attacks from a much more holistic viewpoint.

\printcredits

\bibliographystyle{model1-num-names}

\bibliography{references}

\vskip3pt

\bio{fig/P}
Mohammad Ammar is currently a final year undergraduate at the Dept. of Computer Engineering, Zakir Husain College of Engineering and Technology (ZHCET), Aligarh Muslim University, Aligarh. He is also the Chairperson for the Marine Technology Society Autonomous Underwater Vehicle Club, ZHCET, AMU. His interests lie in Artificial intelligence, Deep-learning, and Underwater Marine Robotics.
\endbio

\bio{fig/IrfanSir}
Dr. Irfan Khan is an Assistant Professor at the Department of Marine Engineering Technology with a courtesy joint appointment with the Electrical and Computer Engineering at Texas A\&M College Station. Before joining TAMU in 2018, he received a Ph.D. in Electrical and Computer Engineering from Carnegie Mellon University USA. He is also the director of the Clean And Resilient Energy Systems (CARES) Lab, which focuses on the cyber security, reliability, and sustainability of cyber-physical systems, including electric energy systems, drones, marine vessels, and biomedical systems. Recently, he has been presented with several prestigious awards and honors, such as the 2021 Jim Leonard Outstanding Member Award from IEEE Region 5, the Gulf Research Program's Early-Career Research Fellowship: Offshore Energy Safety (Track 3) from the National Academies, and the 2021 IEEE Region 5 Director’s Award Technical Conference Co-Chair. 
\endbio

\end{document}